\documentclass[prb,showpacs,twocolumn,floatfix]{revtex4}   %Preprint
\usepackage{graphicx}                             %DVI/PS output
\usepackage{color}
\usepackage{amsmath}
\usepackage{epsfig}
\usepackage{pifont}
\usepackage{hyperref}
\usepackage{pstricks,pst-node,pst-text,pst-3d}
\usepackage{subfigure}
\begin{document}
\def \be {\begin{equation}}
\def \ee {\end{equation}}
\def \nn {\nonumber}
\newcommand{\KF}[1]{\textcolor{red}{\bf [#1]}}
\newcommand{\AS}[1]{\textcolor{blue}{\bf [#1]}}
\title{Screening model of metallic non-ideal contacts at integer quantized Hall regime}
\author{D. Eksi$^1$}
\author{O. Kilicoglu$^1$}
\author{O. G\"oktas$^2$}
\author{A. Siddiki$^{3,4}$}

\affiliation{$^1$Trakya University, Department of Physics, 22030 Edirne, Turkey}
\affiliation{$^2$Braun Center for Submicron Research, Department of Condensed Matter Physics, Weizmann Institute of Science,
Rehovot 76100, Israel}
\affiliation{$^3$Physics
Department, Faculty of Sciences, Istanbul University, 34134-Vezneciler, Istanbul,
Turkey}
\affiliation{$^4$Physics
Department, Harvard University, Cambridge, 02138 MA,
USA}

\begin{abstract}In this work, we calculate the electron and the current density distributions both at the edges and the bulk of a two dimensional electron system, focusing on ideal and non-ideal contacts. A three dimensional Poisson equation is solved self-consistently to obtain the potential profile in the absence of an external magnetic field considering a Hall bar defined both by gates (contacts) and etching (lateral confinement). In the presence of a perpendicular magnetic field, we obtain the spatial distribution of the incompressible strips, taking into account the electron-electron interactions within the Thomas-Fermi approximation. Using a local version of Ohm's law, together with a relevant conductivity model, we also calculate the current distribution. We observe that the incompressible strips can reside either on the edge or at the bulk depending on the field strength. Our numerical results show that, due to a density poor region just in front of the contacts, the incompressible strips do not penetrate to the injection region when considering non-ideal contact configuration. Such a non-ideal contact is in strong contrast with the conventional edge channel pictures, hence has a strong influence on transport. We also take into account heating effects in a phenomenological manner and propose a current  injection mechanism from the compressible regions to the incompressible regions. The model presented here perfectly agrees with the local probe experiments all together with the formation of hot-spots.
\end{abstract}
\pacs{73.23.Ad, 73.43.Fj, 73.43.-f}
\date{\today}
\maketitle

\section{INTRODUCTION}
Regardless of what system one is interested in theoretically~\cite{siddiki2004,Akera06:,Guven03:115327,Chalker07:MZI}, the metallic contacts deposited on the two dimensional charge systems are the most important ingredients of the measurement, both experimentally~\cite{Ahlswede01:562,Ahlswede02:165,Goldman05:155313,Heiblum05:abinter,Goektas:Diss} and theoretically~\cite{Tung,Buettiker86:1761,Tobias:contact,Oswald:contacts}, where by contacts we mean both the injection/collector and probe contacts. In a standard Hall experiment (not necessarily quantized) an external magnetic field $B$ is applied to the system in the direction normal of the plane. The excess electrons are injected from the source contact and are collected at the drain contact by applying a finite potential difference between these two contacts. Meanwhile, one measures the (electrochemical) potential difference either between the contacts along the sample (longitudinal potential difference $V_l$, in the case of a fix imposed current $I$, resistance, $R_l$ measured between A1-A2 or B1-B2 in Fig.~\ref{fig:fig1}) or across the sample, the Hall potential $V_H$ (or resistance $R_H$, measured between A1-B1 or A2-B2 in Fig.~\ref{fig:fig1}). Determining the actual contact resistances of each element is a cumbersome matter. Experimentally, the usual way of obtaining the (non-ideal) contact resistance is the transmission line method, which gives an average approximate value by measuring a set of contacts defined on a stripe of mesa containing 2DES~\cite{Oktay:PhysE:08}. However, in the quantized Hall regime the determination of contact resistances becomes somewhat easier, since one \emph{a priori} knows that the longitudinal resistance should vanish and Hall resistance is quantized to an integer sub-multiple of the von Klitzing constant $R_K=h/e^2=25,812.807449(86)$ $\Omega$, where $e$ is the elementary charge and $h$ is the Planck constant~\cite{vKlitzing80:494}. The term non-ideal contact stands for a case where the edge-channels do not equilibrate at the contacts, hence reflection (or transmission) from channel to contact (or \emph{vice versa}) is non-zero (not unity). In contrary, in an ideal contact \emph{all} the channels are in equilibrium with the contacts, and reflection is zero. We should also clarify the notation \emph{ideal Ohmic contact}: This term stands for a contact with finite resistance $R_C$, however, still obey the Ohm's law $V=IR_C$ where the applied current is $I$. Hence, a non-ideal contact for the Landauer-B\"uttiker formalism, \emph{i.e.} non-zero reflection, can still be an ideal Ohmic contact. A very comprehensive review is provided in Ref.~\cite{Tung}.

It is a formidable task to model the sudden change of the density of states (DOS) near the contacts, while at the metallic region the DOS is approximately infinite, whereas at the 2D system the DOS is either a constant (without magnetic field $B$, $D_0$) or varies between no DOS at the Fermi energy (incompressible~\cite{Chang90:871,Lier94:7757}) and high DOS (compressible, with a degeneracy of $eB/h$), in the presence of a high magnetic field. Recently, there were numerical efforts to model the contacts by graphical methods~\cite{Tobias:contact} considering interacting classical electrons or by the non-equilibrium network model~\cite{Oswald:contacts}. The ideal contacts are modeled as equipotential surfaces, where no density poor region resides just in front of the contacts. Whereas, the non-ideal contacts were described by different injection probabilities at the injection region by Kramer \emph{et al}, however, density fluctuations near the transition region, as indicated by experiments~\cite{Ahlswede02:165,Goektas:Diss} which influences the current distribution drastically is left unresolved.

Here, we present our self-consistently calculated results which takes into account Landau quantization and direct Coulomb interactions at a mean-field approximation level~\cite{Sefa08:prb,SiddikiMarquardt}.

In this paper, we investigate theoretically the current and charge density distributions considering quantizing high $B$ fields and intermediate electron density concentrations in the close proximity of the contacts, together with the entire sample. Our results point that, the assumption of ideal contacts cannot be justified, when considering density inhomogeneities. As a guiding estimate, we employ the relevant formula provided by Chklovskii et al~\cite{Chklovskii92:4026} to calculate the incompressible width just in front of the contact,
\be a_1^2=\frac{16}{\pi}a_B^*l_d[(\nu_0^2-1)/\nu_0^2]\lesssim \lambda_F^2, \ee
where, $a_1$ is the width of first incompressible strip, $a_B^*$ ($\sim 9.81$ nm for GaAs/AlGaAs) is the effective Bohr radius, $l_d$ is the depletion length, $\nu_0$ the bulk filling factor and $\lambda_F$ is the Fermi wavelength. For a typical sample, \emph{i.e.} $\lambda_F \sim 30$ nm, and for $\nu_0 =1.1$ one observes that the depletion length should be less than 20-30 nm (4-5 times the Bohr radius), which is much smaller compared to the experimental findings~\cite{Goektas:Diss}. Hence, models based on ideal contacts has to be revised~\cite{Halperin82:2185,Buettiker88:317}, where the bulk electron density cannot be reach within few $a_B^*$. Moreover, we present clearly the formation of hot-spots~\cite{Ploog:optic} depending on the current amplitude and field direction. We also take into account Joule heating following the lines of H. Akera and his co-workers, phenomenologically and propose a self-consistent mechanism to inject current from compressible to incompressible regions (and \emph{vice versa}). Our treatment of the contact regions is essentially based on the experimental findings of J. Weiss and his co-workers, namely scanning force microscopy experiments by E. Ahlswede~\cite{Ahlswede02:165} and comprehensive structural investigation by O. Goktas~\cite{Goektas:Diss}. Up to our knowledge, no comprehensive study of contacts is accessible in the literature that both take into account interactions, device geometry, potential fluctuations and formation incompressible and compressible regions, which exist due to Landau quantization.
\section{The Modeling of contacts}
\begin{figure}[t]
\centering
\includegraphics[scale=0.20]{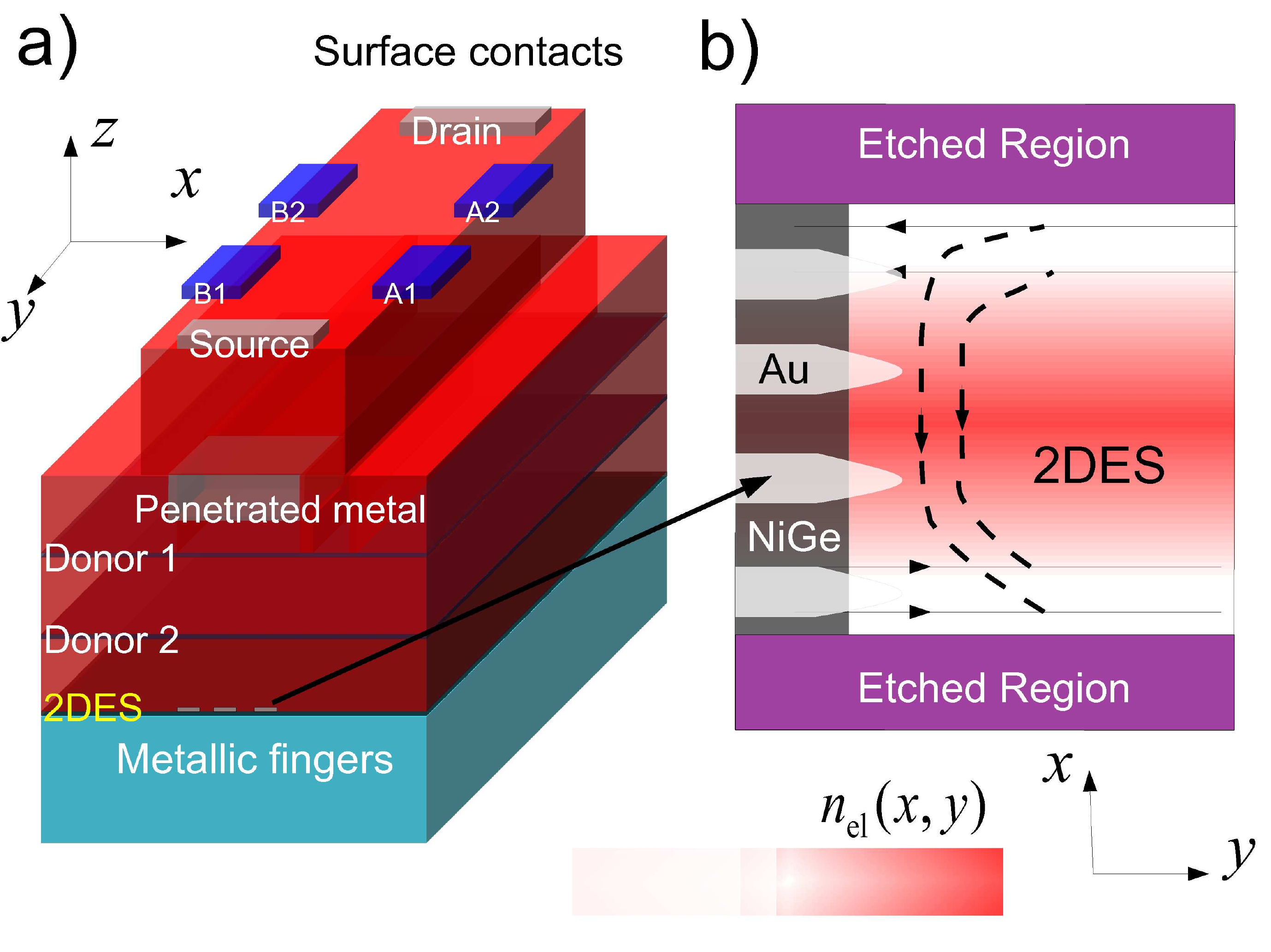}
\caption{(Color online) Schematic presentation of the Hall bar and contacts on a 3D GaAs/AlGaAs (red and blue regions) heterostructure (a). The sample has the dimensions of $1.5\times 1.5 \times 0.85$ $\mu$m$^{3}$, mapped on a matrix of $96\times96\times110$ embedded in a dielectric material, where open boundary conditions are assumed on the border of the dielectric material. Two donor layers, distributed homogeneously on the $xy$ plane, provide electrons both to surface and to the 2DES. Contacts are modeled by i) A metallic gate on the surface ii) A metallic region inside the crystal iii) Fingers at the plane of 2DES, where fingers are to simulate bad contacts. (b) A sketch of 2D projection of the Hall bar from top: Dark (gray and light gray) regions depict the contact, whereas the lateral confinement is shown by the etched regions. Color gradient represents the electron density, together with Landauer-B\"uttiker edge states, in the case of an ideal contact (straight lines) and an non-ideal contact (broken curves).}\label{fig:fig1}
\end{figure}
In spite of the importance of the contacts at 2DESs, it was only recent that systematic investigations revealed the actual material composition at the close vicinity of the contacts~\cite{Goektas:Diss}. They found that, the contacts annealed with the Ni/Ge/Au composition has the following properties: i) The 2DES falls apart from the contacts at the Au-rich regions with an electron depleted region that extends 50 to 150 nm. ii) In contrast, at the close proximity of NiGe-rich grains the 2DES can reach up to the metallic region, however with a density gradient till the bulk density $n_{\rm el}$ is reached. The question how the current is injected to this region is answered by a model deduced from temperature dependent resistance experiments pointing that a very high and thin Schottky-barrier is formed at the contact/GaAs interface, hence the injection is due to tunneling. The most striking result of such an experimental finding is that the B\"uttiker type edge-states cannot equilibrate within the contacts considering an electron poor region just in front of the contacts. For sure, for thin (Schottky) barriers and no poor region LB formalism is adequate and can be extended to thick barrier regime. However, the electron density gradient in front leads to almost full reflection of the B\"uttiker type edge-states, as depicted in Fig.~\ref{fig:fig1}b.

In the above work it was pointed that a model which also takes into account such a density gradient in front of the contacts is admired, especially when considering high perpendicular magnetic fields, where compressible and incompressible strips form. Here, we present such a model by solving the Poisson equation in 3D self-consistently to obtain potential and charge density distributions near the contact, together with the entire sample, employing a forth order grid technique~\cite{Andreas03:potential,Sefa08:prb}. In the next step we use the potential profile obtained at the plane of 2DES as an initial condition to calculate same quantities in the presence of a perpendicular $B$ field. Our calculation scheme is based on a mean-field Thomas Fermi approximation improved by spatial course-graining to simulate quantum mechanical effects, such as the finite extent of the wavefunctions~\cite{siddiki2004}. At a last step we impose a fixed current in $y$ direction and calculate the current distribution utilizing a local version of the Ohm's law~\cite{Guven03:115327}. Here, we assume that the density of states (DOS) and the temperature of electrons do not vary considerably on the quantum mechanical length scales, \emph{i.e.} these quantities are position independent. The conductivity model is obtained from the self-consistent Born approximation~\cite{Ando82:437,siddiki2004} (SCBA) and local conductances are assumed to be directly related with the local electron density. Such an approximation is somewhat crude, however, is valid if the charge density changes slowly on the correlation length of the remote impurities~\cite{SiddikiEPL:09}. An improved calculation scheme~\cite{Champel08:124302}, already supports our assumption on the conductivity model. Moreover, our results are independent of the choice of the conductivity model, since we are not interested in the details of plateau to plateau transition regions. The only necessary ingredient of our conductivity model is to have vanishing (in fact exponentially small, at zero temperature) longitudinal conductivity at incompressible regions and a Hall conductance proportional to the electron density. The last requirement also implies that, the Hall conductance is quantized at the incompressible regions.

In Fig.~\ref{fig:fig1}a we show schematic presentation of the system at hand. The 2DES lies some 150 nm below the surface, where surface potential is pinned to the mid gap of GaAs. Narrow Hall bar is defined by etching at the sides by a depth similar to 65 nm, whereas the metallic fingers reside at the plane of 2DES. To simulate the effects resulting from annealing, we also introduced a metallic region 65 nm below the surface. The contact region is highlighted in Fig.~\ref{fig:fig1}b, where dark regions correspond to gold alloy and light regions represent NiGe alloy. The front of gold alloy is strongly depleted, shown by the white region at the 2DES, whereas, an electron poor stripe resides next to the NiGe alloy. For further references, we define the finger widths ($w_f$) and lengths ($l_f$) differently to simulate good ($w_f=0$ nm and $l_f=0$ nm) and bad contacts ($w_f\gtrsim65$ nm and $l_f\gtrsim190$ nm). In the case of good contacts we assume that the depleted regions in front of the Au grains is negligibly small, whereas for the bad contacts the depleted region is generated with the metallic fingers by keeping the potential at -0.45 V and there are no fingers at the regions corresponding to NiGe grains. The strong confinement in $x$ direction is modified by the contact region.
\begin{figure}[t]
\centering
\includegraphics[scale=0.15]{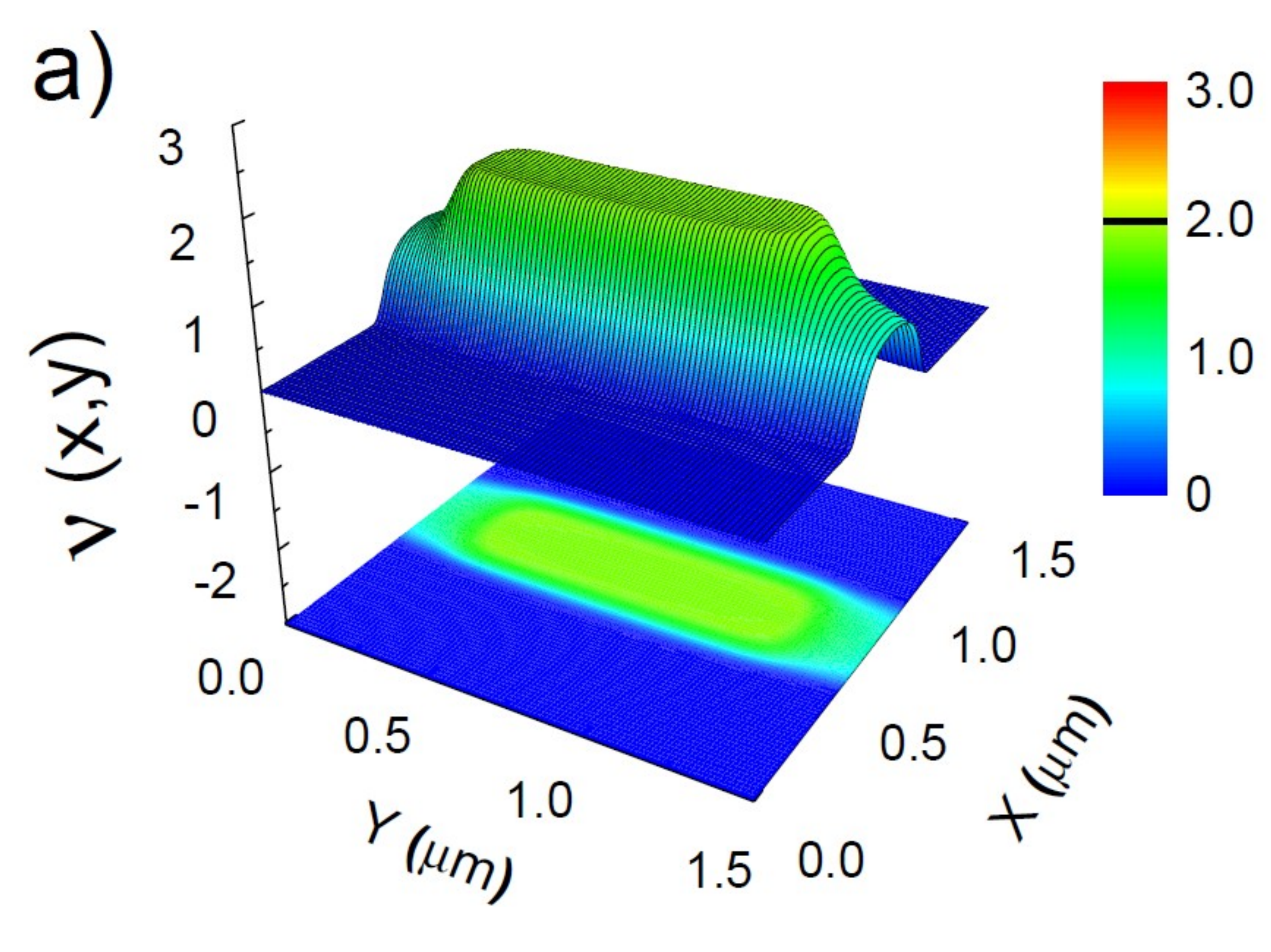}
\includegraphics[scale=0.15]{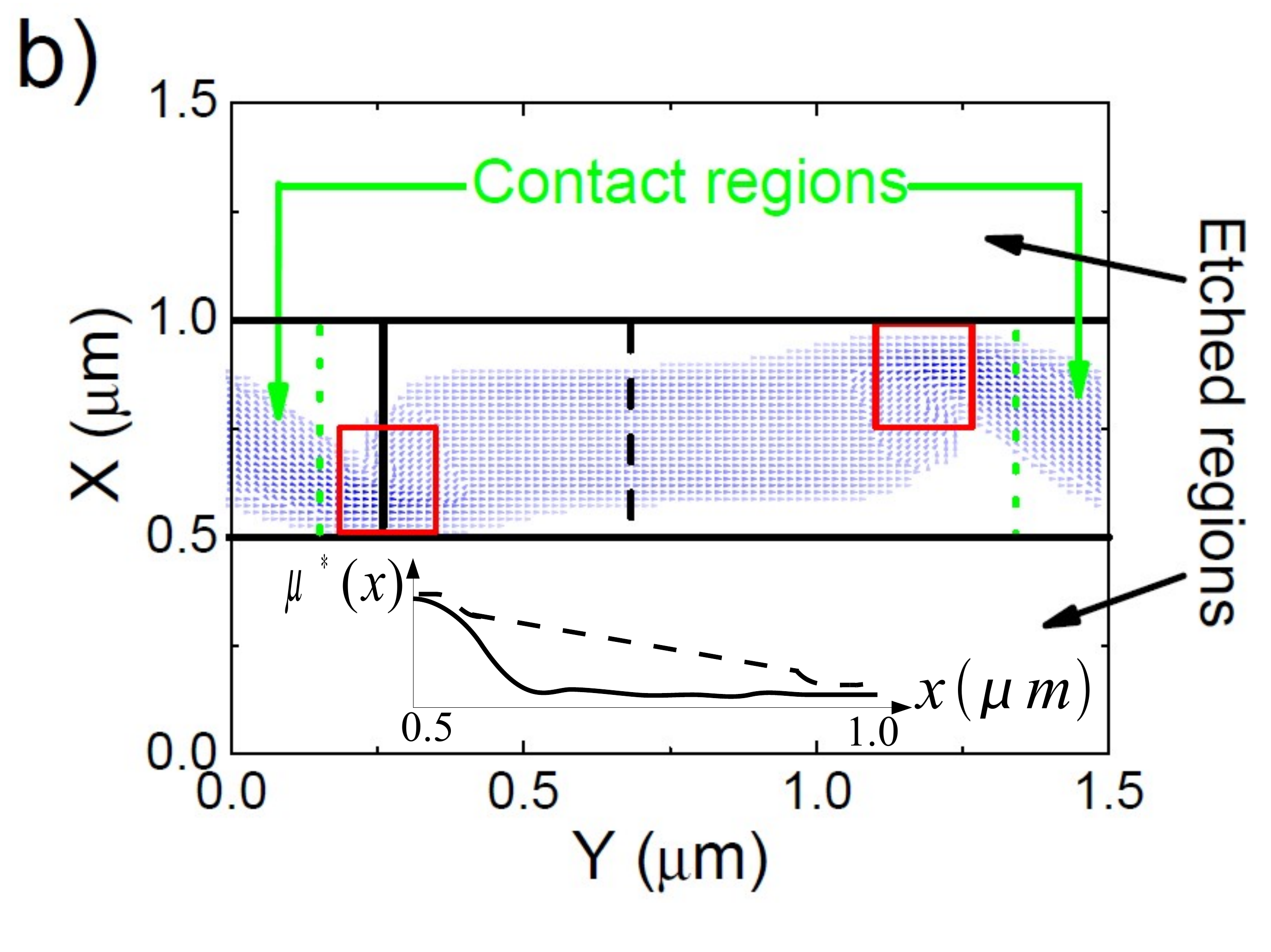}
\caption{(Color online) Spatial distributions of the filling factor $\nu(x,y)$ (a), together with the current density (b) as a function of lateral coordinates. Color scale denotes density gradient, whereas arrows (blue) present the amplitude and direction of the imposed excess current with an amplitude $<0.2$ nA, guaranteeing liner response regime. The calculations are performed at $\hbar\omega_c/E_F=4.38$ and default temperature $kT/E_F\sim0.08$. The electron poor region is a result of metallic contacts at the surface and penetrated metal, kept at -0.5 V. The associated electrochemical potential distributions are shown in the inset of (b), following the vertical solid and broken lines, respectively. Boxes indicate the hot-spot regions, separated from the contact regions by broken (green) vertical lines. Here, contact region is defined by the density gradient, hence Landau levels also do form in this region. The current is injected exactly from $y=0$ $\mu$m line.}\label{fig:fig2}
\end{figure}
We begin our discussion with a case which has an one to one correspondence in \emph{classical} Hall effect: We set the $B$ field such that only the lowest Landau level is partially occupied, hence, behaves like a metal. Throughout this paper, following the previous works, we neglect the spin degeneracy, since the essential physics is independent of the origin of the single particle gap, however, spin generalized versions of the screening theory already exist~\cite{afifPHYSEspin,GonulTFD:09}. It is common to describe the 2D electronic system in the presence of a perpendicular field  by a dimensionless parameter; the filling factor, $\nu$, which measures the occupancy of the Landau levels below the Fermi energy given by the ratio of electron number density $n_{\rm el}$ to the magnetic flux density $n_{\phi}$. One can express the filling factor also in terms of magnetic length $l_B$ $(=\sqrt{\hbar/eB})$ as $\nu=2 \pi l_B^2n_{\rm el}$ and the local version as $\nu(x,y)=2 \pi l_B^2n_{\rm el}(x,y)$. Therefore in this case the filling factor is below 2 (due to spin degeneracy) and there are many available states at the Fermi energy, similar to a metal. Such a case (also locally) is called as compressible, where screening is nearly perfect and electrons can be redistributed according to applied electric fields. Fig.~\ref{fig:fig2}a presents the filling factor distribution (or electron density) calculated for the geometry described above, however, without metallic fingers at the plane of 2DES to mimic an ideal \emph{Ohmic} contact. Due to the strong confinement by the etching, a large electron depleted region resides on both sides of the Hall bar. Note once more that, such a contact with a density poor region is not an ideal contact considering Landauer-B\"uttiker edge channel picture, as depicted in Fig.~\ref{fig:fig1}b. One observes that, the electronic density is slightly small in front the contacts compared to the bulk. If we consider the classical Hall effect, the current (and electrochemical potential) distribution is well known~\cite{Asch:book}: there will be two spots where electrochemical potential is highest (depending on the field direction) at the right-bottom and lowest at the left-top corners. For an illustrative demonstration we suggest the reader to check Fig.1 of Ref.~\onlinecite{Tobias:contact} and also Fig.4 to compare with actual first principle many-body simulations, performed at the classical regime together with interactions. The experimental findings also support the formation of these spots~\cite{Ploog:optic,Ploog96:289,Ahlswede01:562}. Here the current amplitude is measured in units of $e v_F k_F/(\pi L_y)$, where $v_F$, $k_F$ are the Fermi velocity and wave vector, respectively, and $L_y$ $(=1500$ nm) is the length of the sample, hence the current amplitude is less than 0.2 nA directed in positive $y$ direction. The current distribution also presents two regions where most of the transport takes place, denoted by (red) boxes in Fig.~\ref{fig:fig2}b, we observe that the current is almost homogeneously injected by the source contact, however, is confined to these regions. Once the current enters to the bulk of the sample, it is distributed homogeneously following the local electron density, likewise the classical Drude result~\cite{Asch:book,Siddiki04:condmat}. This regime overlaps perfectly with the local probe experiments~\cite{Ahlswede02:165}, where we also show the demonstrations of the electrochemical potential distribution across the sample in the lower inset of Fig.~\ref{fig:fig2}b.

\begin{figure}[t]
\centering
\includegraphics[scale=0.15]{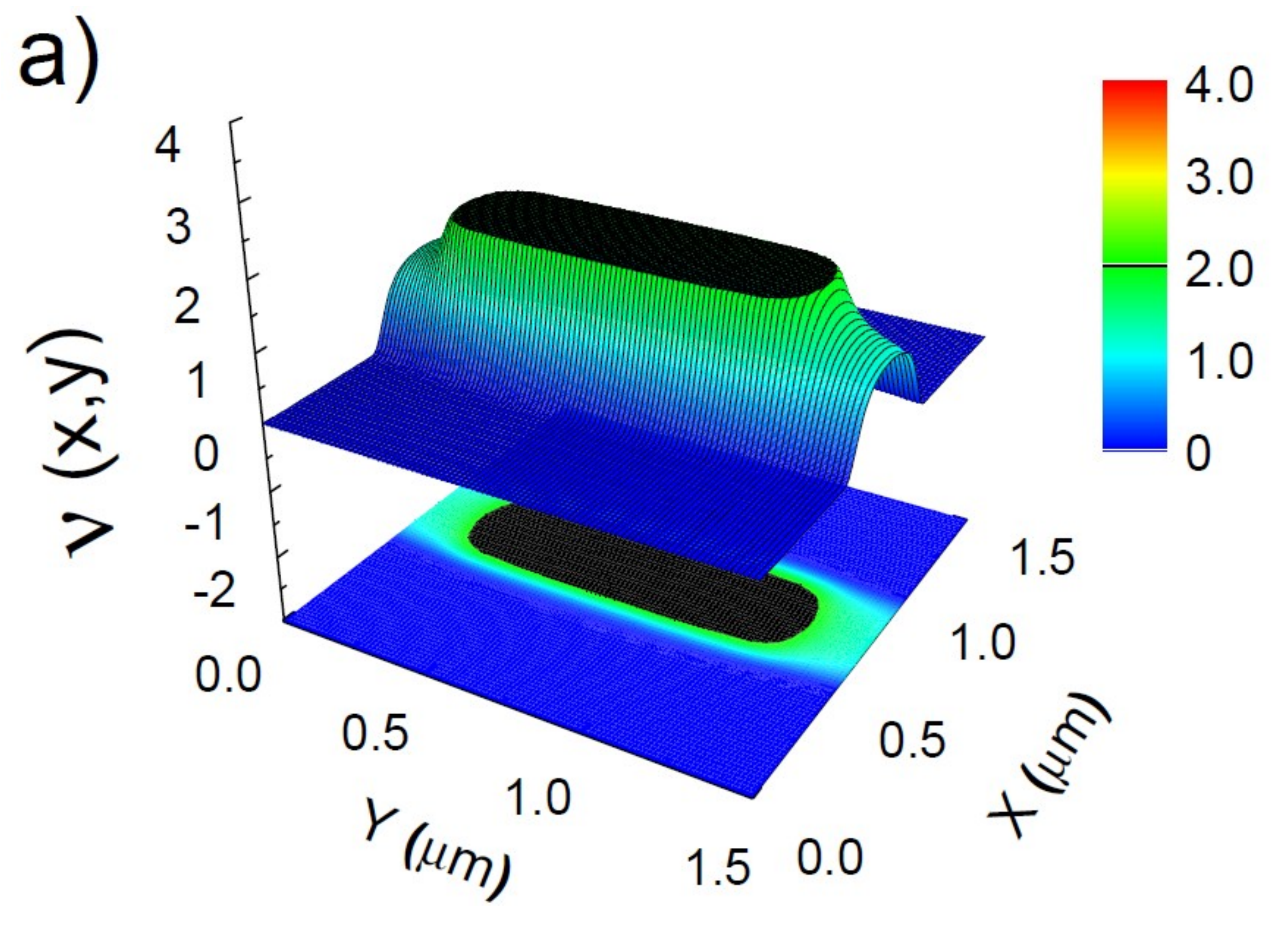}
\includegraphics[scale=0.15]{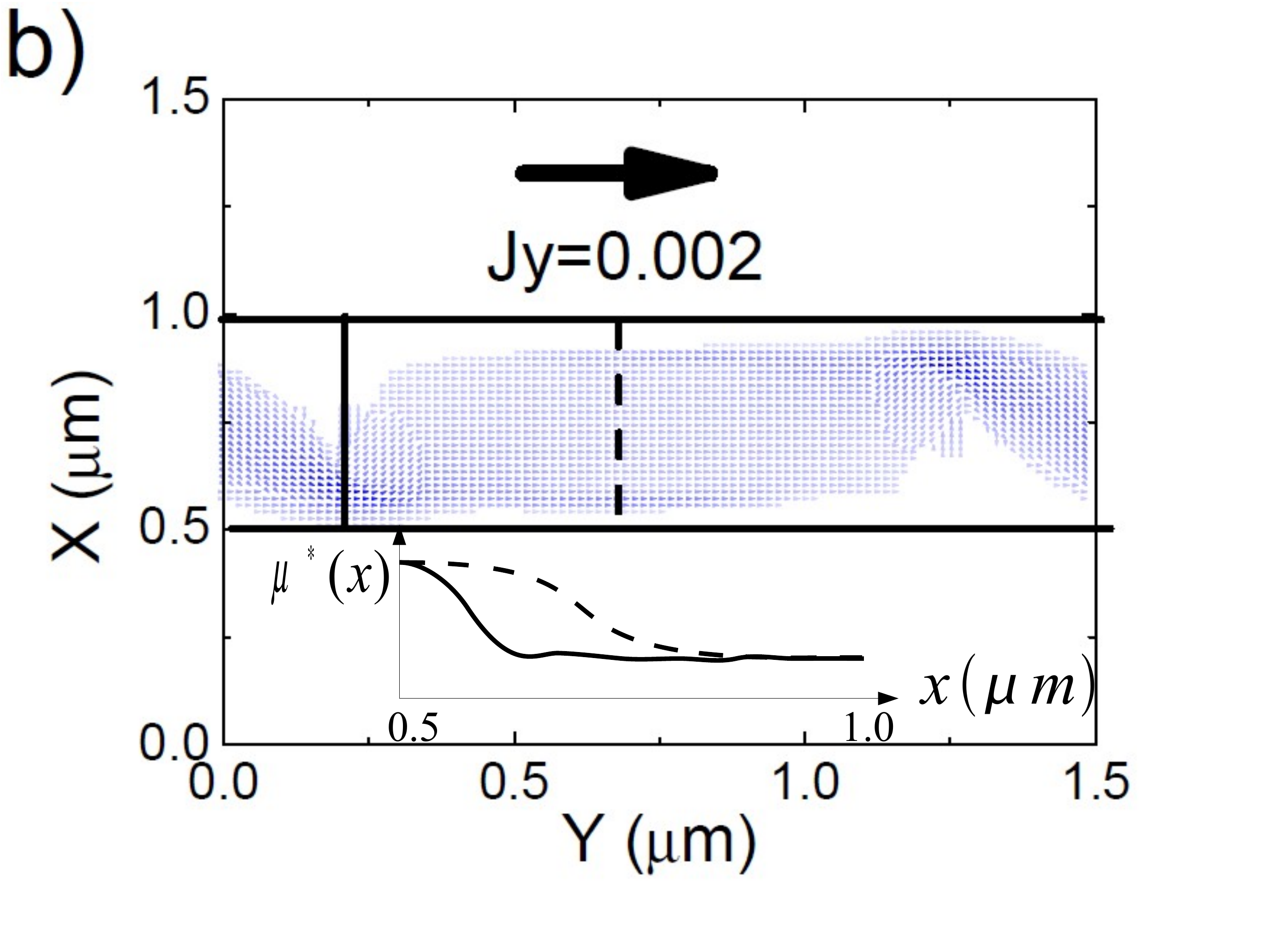}
\includegraphics[scale=0.15]{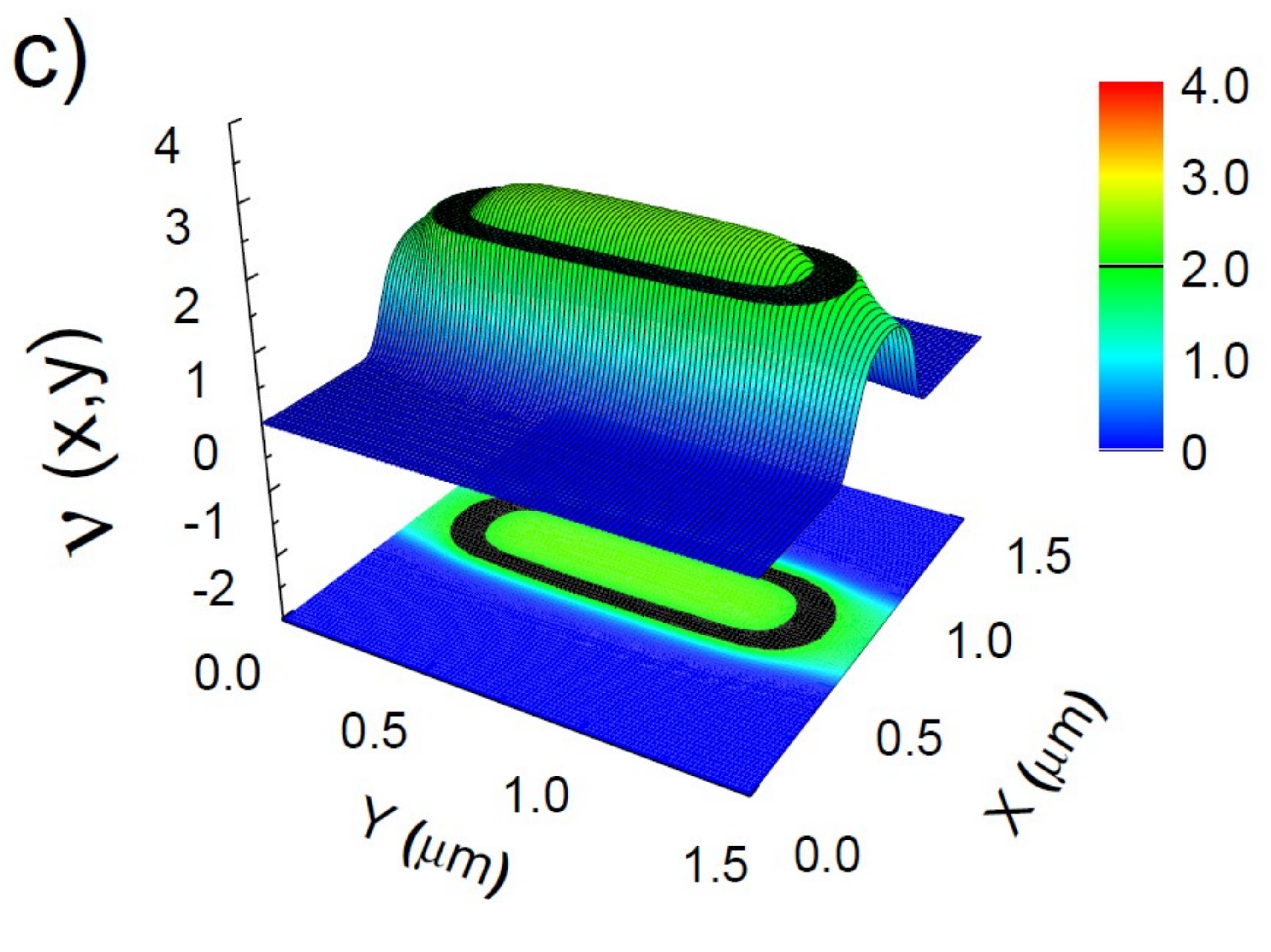}
\includegraphics[scale=0.15]{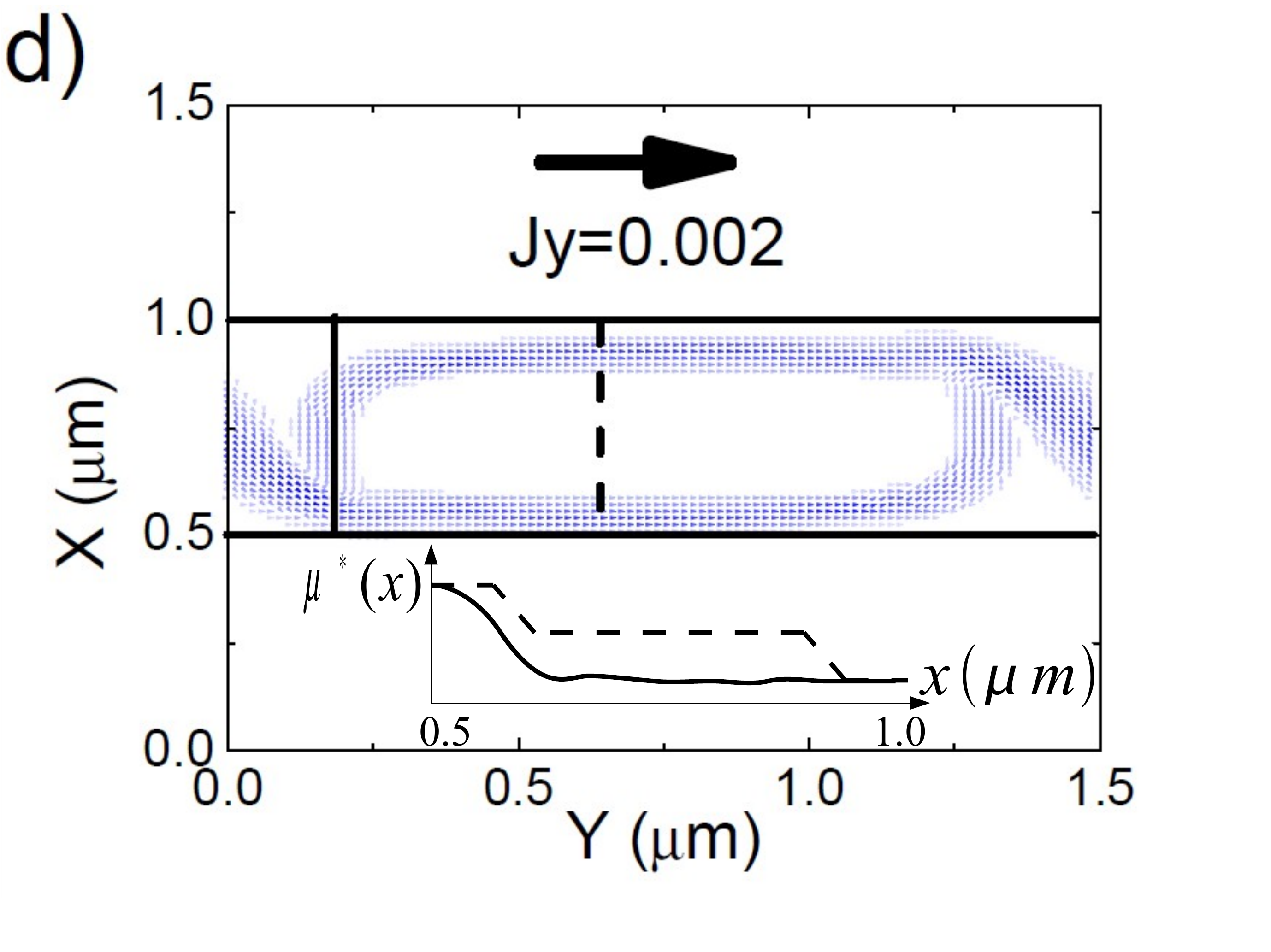}
\includegraphics[scale=0.15]{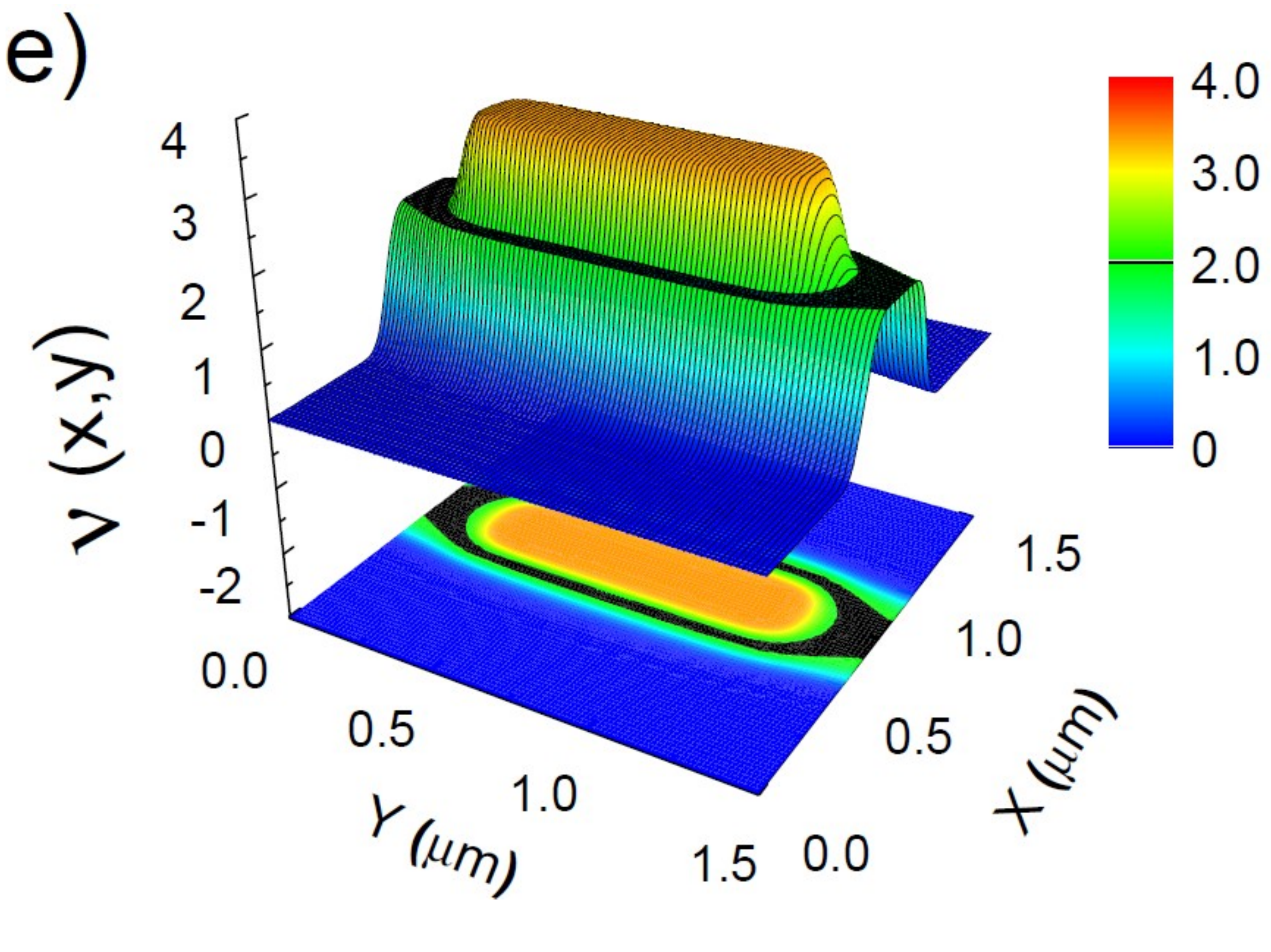}
\includegraphics[scale=0.15]{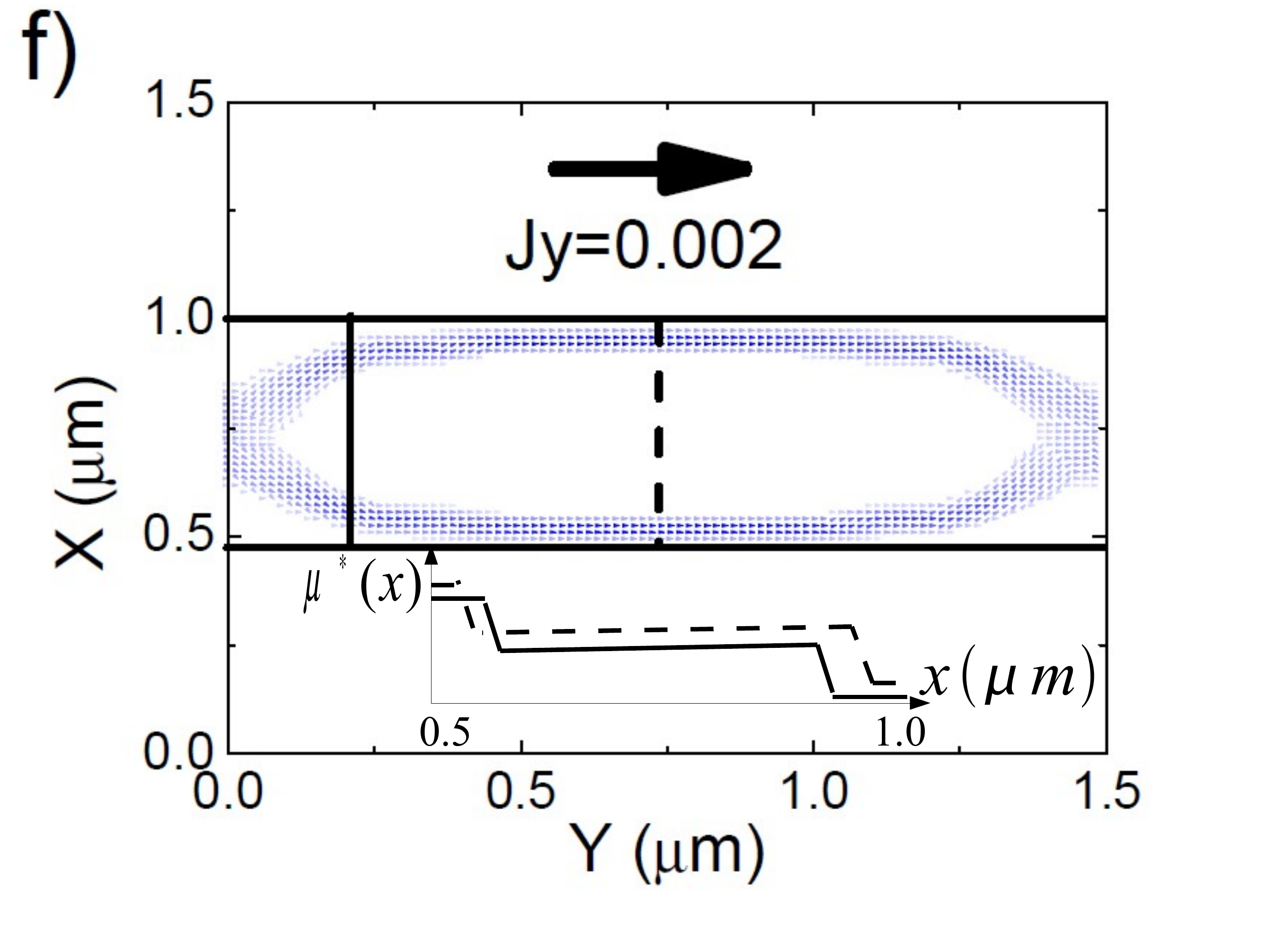}
\caption{(Color online) Spatial distribution of the electron densities (left panel) and current densities, same as Fig.~\ref{fig:fig2}, calculated at selected $B$ field strengths, $\hbar\omega_c/E_F=$ 3.78 (a-b), 3.02 (c-d) and 2.26 (e-f) at default temperature. (e-f) mimics an ideal contact in edge-channel picture, where no hot-spots are observed. Insets at the right panel demonstrate the electrochemical potential distributions similar to the previous figure.}\label{fig:fig3}
\end{figure}

Next, we lower the $B$ field such that a wide incompressible region (denoted by black) is formed at the bulk of the sample, Fig.~\ref{fig:fig3}a. We also show the corresponding current distribution in Fig.~\ref{fig:fig3}b, where a positive current is imposed in $y$ direction. The hot-spots are still visible, however, now the current is injected to the incompressible bulk instead of a compressible bulk. Injecting current to an \emph{incompressible} region is somewhat counter intuitive at a first superficial look. Before elucidating this, we would like to emphasize that our results presented are calculated self-consistently for a given background potential and imposed current, \emph{i.e.} finite electrochemical potential difference between the source and drain contacts, together with the local Ohm's law. Moreover, we do not \emph{assume} that the current should be injected to the incompressible regions. However, self-consistency results in such a case. In fact the key concept is the electrochemical potential difference between contacts: We inject electrons which have an electrochemical potential energy larger than the Fermi energy otherwise (namely, if they were all at the Fermi energy) there would be no net current. It is useful to re-introduce~\cite{Guven03:115327} the definition of (local) Ohm's law;
\be \nabla\mu^*(\textbf{r})/e=\hat{\rho}(\textbf{r})\textbf{j}(\textbf{r})=\textbf{E}(\textbf{r}), \ee
where $\mu^*(\textbf{r})$ is the position dependent electrochemical potential, $\hat{\rho}(\textbf{r})$ is a two by two tensor describing the local resistivities and $\textbf{j}(\textbf{r})$ is the local current density together with the local electric field $\textbf{E}(\textbf{r})$. The calculation procedure is such: First we obtain the electrostatic properties of the system at total equilibrium, (\emph{i.e.} $\mu^*$ is position independent and equals to $E_F$), then impose a small (compared to equilibrium) electrochemical potential difference between source and drain contacts and re-calculate the electron distribution depending on newly calculated total potential. Once local equilibrium conditions are satisfied together with the numerical convergence, namely the potential and electron density distributions do not change within the numerical accuracy of $10^{-7}$ in the last iteration, the imposed electrochemical potential difference is increased step by step to the target value. In fact, in our calculations presented here, we imposed a very small current amplitude therefore, the density distribution does not change substantially. In other words, we are in the linear response regime, however, we are not limited to this regime and our results in the non-linear regime are published elsewhere~\cite{denizphyEvelocity,Sefa08:prb}. Moreover, if one takes into account heating effects~\cite{Akera06:} due to the high current density at the hot-spots, it is easy to see that the incompressible strips at the injection and emission regions are \emph{melted}, hence the excess current can be transferred between compressible and incompressible regions (and \emph{vice versa}) easily. Including such local temperature effects are far complicated to be incorporated within our calculation scheme. However, one can still estimate the local electron temperature variation at the hot-spots from the work by H. Akera, namely Ref.~\onlinecite{Akera06:}, focusing on Fig.1 and Fig.2b (as a function of filling factor and lattice temperature) or as a function of imposed current amplitude (Fig. 7). Depending on the parameters we employed here, we estimate that, the electron temperature at the hot-spots are 5-10 (and even more in strip case) percent higher than the lattice temperature, which essentially implies that the incompressible region at the hot-spot melts, \emph{i.e} the single particle energy gap closes due to large derivative of the Fermi function. In the inset of Fig.~\ref{fig:fig3}b, we also depict the spatial variation of the associated electrochemical potential. It is observed that, at the hot spot region $\mu^*(x)$ varies in a highly nonlinear manner, having a maximal variation exactly at the spot (solid thick line), whereas, far from the injection contact potential presents an s-shape behavior (broken line) similar to the experimental~\cite{Ahlswede01:562} and theoretical~\cite{Siddiki04:condmat} findings. Hence, the Hall resistance measured at the center is quantized, however, near the hot-spots deviate from the plateau value considerably.
\begin{figure}[t]
\centering
\includegraphics[scale=0.15]{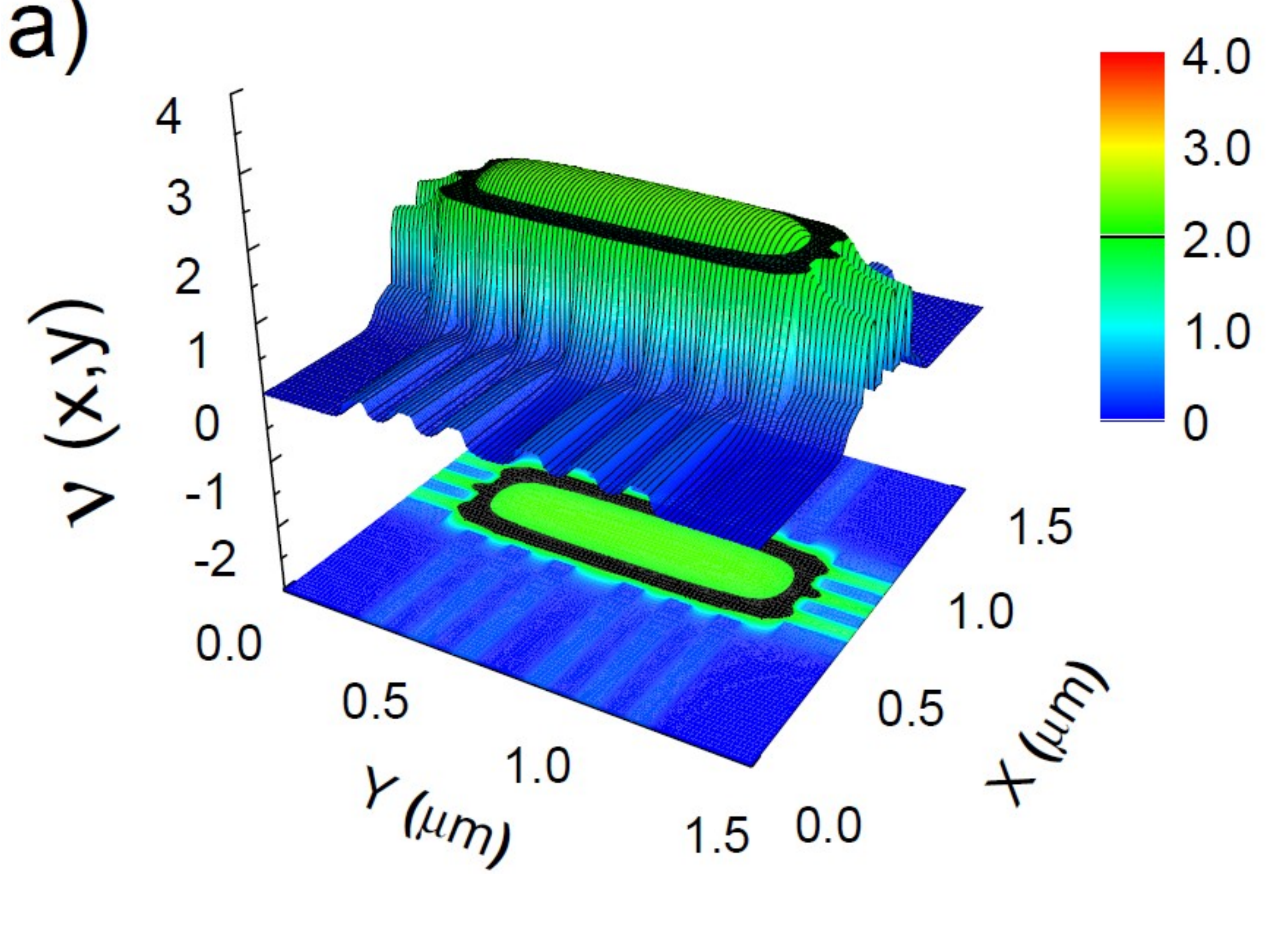}
\includegraphics[scale=0.15]{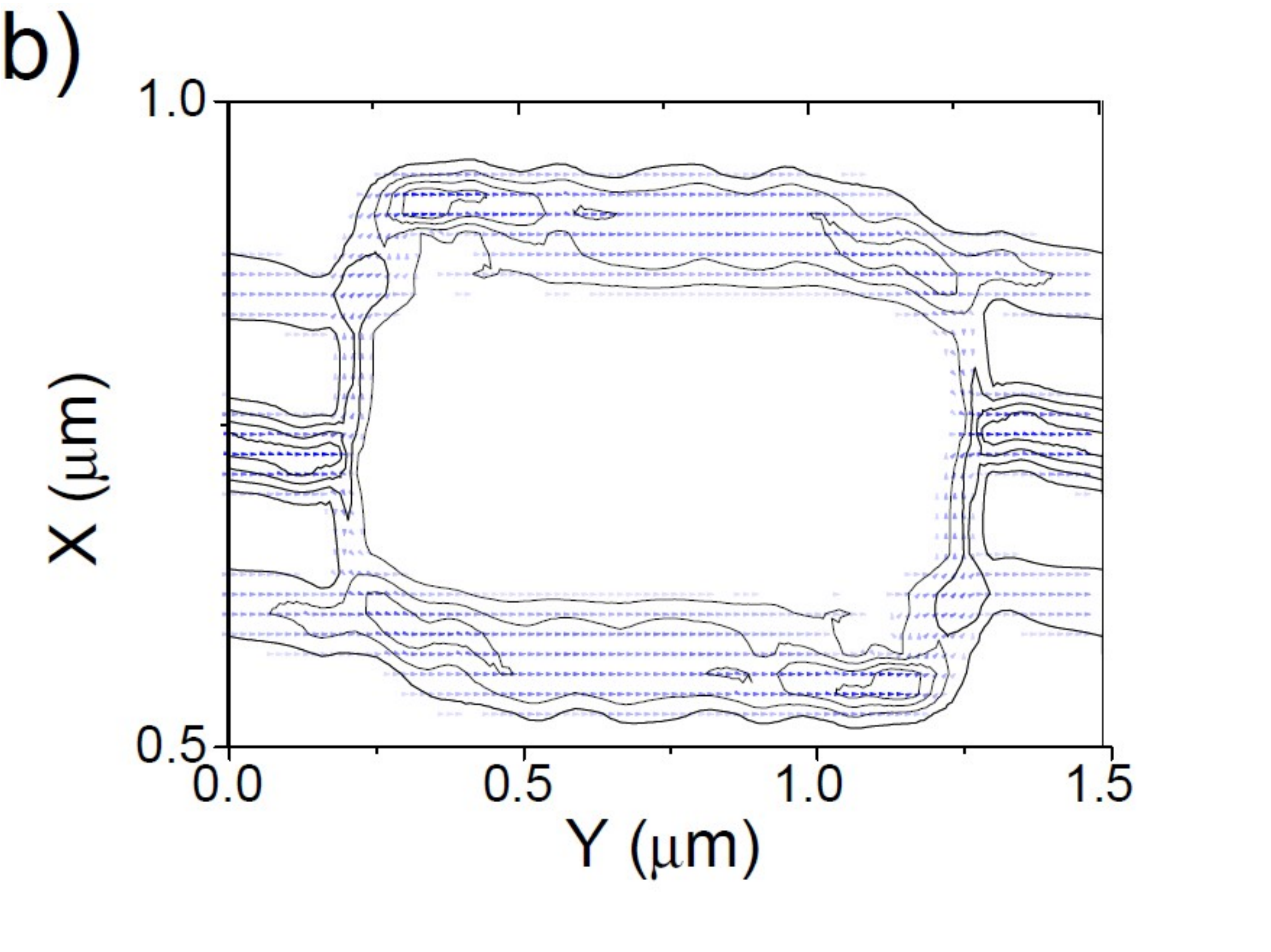}
\caption{(Color online) Self-consistently calculated carrier density (a) and current density distributions (b), where (b) focuses on the 2DES region. The fluctuations at the contour plot present the anomalies near the contacts, smearing out the quantized Hall effect.}\label{fig:fig4}
\end{figure}
Fig.~\ref{fig:fig3}c and Fig.~\ref{fig:fig3}d, corresponds to a magnetic field strength where a deformed incompressible ring is formed close to the edges. Nothing different occurs at injector (and collector) region at this $B$ value, the non-linear potential drop is observed with slight modifications. In contrast the electrochemical potential profile changes substantially at the center, the strong variation only takes place where the incompressible strips reside, inset Fig.~\ref{fig:fig3}e. We observe that, once the current is injected to the incompressible strip it is kept being confined to this region due to the absence of back-scattering, till the next hot-spot. One deduces from the electrochemical potential profile that the $R_H$ is quantized, similar to our previous works~\cite{SiddikiEPL:09} and experiments~\cite{Ahlswede01:562}.

So far we have observed the features of an ideal Ohmic contact (non-ideal in the language of Landauer-B\"uttiker) due to the electron poor region in the front of contacts. However, our results show that the current can be directly injected to the incompressible strips through the compressible region at the contacts. We have seen that, near the contact regions the essential futures of the quantized Hall effect is lost. In contrast far from the contacts it is recovered, we think that these numerical results are in accord with the ones of Ref.~\onlinecite{Oswald:contacts} and with the experimental findings~\cite{Ahlswede01:562,Goektas:Diss}. Now the question is, what happens if an incompressible edge strip touches the contact region, \emph{i.e} an ideal contact. Such a case occurs as a natural result of the self-consistency, which we present in Fig.~\ref{fig:fig3}e. The direct Coulomb interaction and Landau quantization (and Lorentz force), pins the electron density just in front of the contact to an (even) integer filling factor at $ 2.26 \hbar\omega_c/E_F^0$. The corresponding, current distribution is shown in Fig.~\ref{fig:fig3}f, where one cannot observe the formation of the hot-spots. Since, all the current is directed along the incompressible strip starting from the injection contact (namely, due to the absence of back-scattering). In this case, one can think of an ideal contact in the terminology of Landauer-B\"uttiker edge-state formalism. Unfortunately, such a case is in contrast to the local probe experiments~\cite{Ahlswede01:562} and other experiments which report hot-spots~\cite{Ploog:optic}.

The influence of the contact quality on the quantized Hall effect was also investigated in Ref.~\onlinecite{Goektas:Diss}. It was reported that, measurements performed on deeper lying heterostructures considering contacts which are defined in bad contacting direction, the plateaus are not well developed. In this last part we model \emph{bad contacts}, yet utilizing thick metal fingers residing in the plane of the 2DES, namely we set the finger length to be $190$ nm and width $65$ nm, at the injection and collector channels. These metallic fingers essentially represent electronically depleted regions in front of Au grains. Whereas, the Ni/Ge grains correspond to regions in between the fingers (kept at ground), where a density gradient occurs. A typical result is shown in Fig.~\ref{fig:fig4}, considering same parameters chosen in Fig.~\ref{fig:fig3}e and Fig.~\ref{fig:fig3}f, together with contour lines of the current density. Here, we superimposed density fluctuations induced by negatively charged (-0.5 V) metallic fingers to simulate effects resulting from large Au grains, both in the longitudinal (corresponding to injection and collector contacts) and lateral (simulating probe contacts) directions, where latter is defined by thicker fingers ($l_f=530$ nm and $w_f=80$ nm). The first observation is that, the current is no longer injected (almost) homogeneously from the contact to the 2DES, moreover, no incompressible strips form at the inner parts of the fingers due to the strong potential variation. Hence, scattering is enhanced. To simulate the effect of bad side contacts we also impose metallic fingers on the sides. We set the widths and lengths similar to the injection/collector contacts, whereas in Fig.~\ref{fig:fig2} and Fig.~\ref{fig:fig3} these lengths were set to zero. We observe that, the density in the lateral direction also presents oscillations and the widths of the incompressible strips vary considerably, even they become narrower than the extent of the wavefunction. Hence, the back-scattering free strips are lost, resulting in non-quantized Hall plateaus. Instead of hot-spots, we see that the current is diverted to the right-bottom corner due to Lorentz force. The influence of bad probe contacts would be suppressed if these side contacts reside far apart. Unfortunately, our numerical abilities are limited by the dimension of the matrix mapping the Hall bar, therefore we cannot perform systematic investigations in this regime. In any case, our results suggest that, once the density and potential fluctuations imposed by probe contacts are suppressed at large distances the ideal behavior would be recovered. Another important point to note is, once the sample becomes wider and the effect of potential fluctuations are suppressed by screening near the side contacts, the quantized Hall effect would be recovered, since one only measures the (electrochemical) potential difference between the side contacts.
\section{conclusions}
Here, we investigated the influences of contacts on the current and density distributions, considering a Hall bar under quantized Hall conditions, within the screening theory. We modeled the contacts following the ideas put forward by O. G\"oktas based on the experimental findings, namely the formation of an electron poor region in the close proximity of contacts. It is shown that, in the \emph{classical} regime where all the system is compressible and behaves like a metal, the well known hot-spots are formed at the corners of the sample. At lower fields, where Landau quantization becomes important, incompressible strips (or regions) are formed and current is injected to these states via hot-spots in the case of ideal Ohmic contacts. Here we also discussed heating effects in a phenomenological manner and argued that due to high current densities at the hot-spots the incompressible regions melt locally. Moreover, we showed that due to the electron poor region, Landauer-B\"uttiker edge-channels are reflected from the contacts and can be considered as non-ideal contacts. We also provide numerical results such that if the $B$ field and the contact structure is chosen appropriately, one can still obtain an ideal contact in the terminology of Landauer-B\"uttiker formalism. However, one cannot observe the formation of hot-spots in this case which is in contradiction with the experiments mentioned, unfortunately. In a final discussion, we also simulated bad contacts taking into account potential fluctuations resulting from Au/Ni/Ge alloys by placing metallic fingers in the plane of the 2DES. We observed that, the incompressible strips are destroyed and large amount of scattering takes place both near the injection/collector and side probe contacts, hence the quantized Hall effect is lost. Our findings, we think, are in accord with the experimental findings and also with the numerical investigations considering the classical regime and non-equilibrium network model. The investigation of actual injection of electrons through the Schottky barrier and local temperature effects demands a more complicated calculation than presented in this work, however, our numerical investigations support the idea that such an investigation might modify the picture presented slightly in a small parameter space.

While submitting our manuscript, we have encountered a very recent experimental paper reporting on the anisotropic depletion at contact interfaces measured by scanning force microscope~\cite{Dahlem:10}. The reported potential profiles strongly support our model, which identifies presence of an incompressible strip in front of ohmic contacts, that might decouple compressible bulk from the contacts. Moreover, our modeling of good/bad contacts also perfectly agree with their findings.

We thank to E. Ahlswede, F. Dahlem and J. Weiss for introducing us and critical discussions on the scanning force microscope results. A. Weichselbaum is acknowledged for sharing his experiences and 3D Poisson solver routine. Fruitful discussions with T. Kramer is also appreciated. We would like to express our gratitude to R. R. Gerhartds, who introduced us most of the concepts of the screening theory. AS acknowledges E. J. Heller for giving the opportunity to work in his distinguished group at Harvard University, physics department. This work is financially supported by the scientific and research council of Turkey, under grant no: TBAG-109T083.
%\bibliography{zitate}

\end{document}